\def\year{2016}\relax
\documentclass[letterpaper]{article}

\usepackage{aaai16}
\usepackage{times}
\usepackage{helvet}
\usepackage{courier}
\usepackage{graphicx}
\usepackage{amsfonts}
\usepackage{amsmath}
\usepackage{booktabs}
\usepackage{subfigure}
\usepackage{color}
\usepackage{hyperref}

\begin{document}
 
\title{The Road to Popularity: The Dilution of Growing Audience on Twitter}
\author{
Przemyslaw A. Grabowicz\\
MPI-SWS, Germany
\And
Mahmoudreza Babaei\\
MPI-SWS, Germany
\And
Juhi Kulshrestha\\
MPI-SWS, Germany
\AND
Ingmar Weber\\
Qatar Computing Research Institute, HBKU, Qatar
}
\maketitle

\begin{abstract}
On social media platforms, like Twitter, users are often interested in gaining more influence and popularity by growing their set of followers, aka their \textit{audience}. Several studies have described the properties of users on Twitter based on static snapshots of their follower network. Other studies have analyzed the general process of link formation. Here, rather than investigating the dynamics of this process itself, we study how the characteristics of the audience and follower links change as the audience of a user grows in size on the road to user's popularity.

To begin with, we find that the early followers tend to be more elite users than the late followers, i.e., they are more likely to have verified and expert accounts. Moreover, the early followers are significantly more similar to the person that they follow than the late followers. Namely, they are more likely to share time zone, language, and topics of interests with the followed user. To some extent, these phenomena are related with the growth of Twitter itself, wherein the early followers tend to be the early adopters of Twitter, while the late followers are late adopters. We isolate, however, the effect of the growth of audiences consisting of followers from the growth of Twitter's user base itself. Finally, we measure the engagement of such audiences with the content of the followed user, by measuring the probability that an early or late follower becomes a retweeter.

\end{abstract}

%%%%%%%%%%%%%%%%%%%%%%%%%%%%%%%%%%%%%%%%%%
\section{Introduction}
\label{sec:intro}

% illustration
Politicians are sometimes accused of abandoning principles from their earlier days as they rise to power. However, one might argue that they simply adapt their message to be in line with their growing -- and evolving -- target audience. Similarly, as a once unknown actor reaches stardom, their audience changes from a few ``niche'' fans and personal friends, to millions worldwide. In this paper, we study how an average Twitter user's audience, and her engagement with this audience, changes as she accumulates followers.

We start from an observation by Chen et al.\ that ``the pope is not a scaled up bishop'' \cite{chenetal14socinfo}. They observed that same-religion linkage preference on Twitter is lower for users with lots of followers, such as the pope, than it is for less visible users. This led us to hypothesize that as a user grows her audience, the level of topical interest and engagement of the audience decays -- or as we will say ``dilutes''. Understanding if such an effect exists has important consequences, because general alignment between audience and message is fundamental for successful communication~\cite{Grabowicz2016Distinguishing}, not only on Twitter.

For instance, changing properties of a growing audience may have an influence on information diffusion. Previous studies show that pieces of topical information spread more readily among users interested in the topic of such memes~\cite{Grabowicz2016Distinguishing} and that the retweeting behavior is influenced by the posting time of information~\cite{Yang2010Understanding}. 
These results suggest that information diffusion is less efficient if a growing audience becomes less topically aligned with the messages or less synchronized in time. 
% not sure what was the idea behind this sentence
%Also, changes in audience composition will likely have an effect on content recommendation. In fact, user-level evolution of online reviews has an effect on content recommendation~\cite{McAuley2013From}.

Studying the change in audience composition and its dilution for a given Twitter user is challenging, as many confounding factors have to be taken into account. First of all, a random new Twitter user today will be different from a random new Twitter user eight years ago. In fact, Liu et al.~\cite{Liu2014Tweets} show that the distribution of location and language of new Twitter users changes over time. Second, existing users also continue to evolve. Garcia et al.~\cite{garciagavilanesetal14socinfo} observe that ``mature users evolve to adopt Twitter as a news media rather than a social network''. Third, the links in the social network evolve as the network grows~\cite{Leskovec2008Microscopic}. Though an evolving network will surely effect information diffusion, it is less clear how information diffusion relates to the changes in audience composition.

Here, we distinguish between two different hypotheses. Namely, we hypothesize that an audience, as it grows, it dilutes its properties, such as: elitness or the coherence of time zone, language, and topics of expertise. We define as an \textit{audience}: (i) Twitter's internal audiences, where each audience is defined as all followers of an expert, and (ii) the user base of Twitter itself. We would like to learn if any of these audiences or both of them dilute their properties as they grow.

In our analysis, we isolate the effects induced by the dilution of a growing audience of a particular expert user and the dilution of Twitter's growing user base itself. To this end, as a control variable we use the time when a given follower joined Twitter, i.e., the \textit{age} of that user in the system. This allows us to measure both effects of the growing number of followers and Twitter's user base on the composition of such audiences.
We find that:
\begin{itemize}
\item As the number of followers grows for a given expert user, the composition of these followers dilutes. We find examples of this rule in the eliteness of followers and their topical coherence.
\item The composition of Twitter's user base also dilutes. Namely, the eliteness of users joining Twitter drops and so does their time zone and language coherence.
\end{itemize}
In what follows, we first describe the dataset that illustrates our findings. Then, we introduce the random shuffling method that is used to distinguish between the growth of the number of followers and Twitter's user base (Section~\ref{sec:eliteness}). Next, we measure the properties of these growing audiences (Section~\ref{sec:analysis}) and their engagement estimated as the fraction of followers diffusing information (Section~\ref{sec:infodiff}).

%%%%%%%%%%%%%%%%%%%%%%%%%%%%%%%%%%%%%%%%%%
\section{Description of the dataset}
\label{sec:dataset}

% sampling
In this work, we analyze followers of \textit{expert} users. 
%The detailed definition of an expert was introduced in the study~\cite{Ghosh2012Cognos:}. 
We define an expert as a user that was included in multiple Twitter lists~\cite{Ghosh2012Cognos:}.\footnote{A list is a set of users tagged by the creator of the list. See \text{https://blog.twitter.com/2009/theres-a-list-for-that} for details.}
% topics
For each of the expert users, we know the topics of their expertise as they have been explicitly ``tagged'' by the list curators. Also, we represent expertise of each user as a normalized topical distribution over her topics of expertise~\cite{icwsm15kulshrestha}.
Expert users tend to produce more high-quality content and less spam and they are more engaged than random Twitter users~\cite{Ghosh2013sampling}. We choose to focus on random expert users instead of all users, due to these qualities.

In total, we identified around $1.6$M experts in Twitter as of November 2015. Note that users with many followers are likely to be included in this set.\footnote{For comparison, Followerwonk listed 824k Twitter users with at least 10,000 followers as of 5th of January 2016, see \text{https://moz.com/followerwonk/bio/?flmin=10000}.}
For the purpose of this study, we sample a set of $1$K random expert users. 
More specifically, we select random experts from five different bins of the number of followers. 
Namely, we select $200$ experts that each had $100$ to $1$K, $1$K-$10$K, $10$K-$100$K, $100$K-$1$M, or $1$M-$10$M followers in the end of November 2015.

% data downloading
Then, using Twitter's REST API, we retrieve all followers of our expert users and the profile information of these followers. In total, we gather information about $241$M users.
Furthermore, we collect all tweets of our expert users that were posted during three months between September and October 2015 and all retweets of these tweets, amounting to $1$M tweets and $23$M retweets.

%%%%%%%%%%%%%%%%%%%%%%%%%%%%%%%%%%%%%%%%%%
\section{The dilution of growing audience}
\label{sec:analysis}

\begin{figure}[t]
\centering
\subfigure[Verified]{\includegraphics[width=0.22\textwidth]{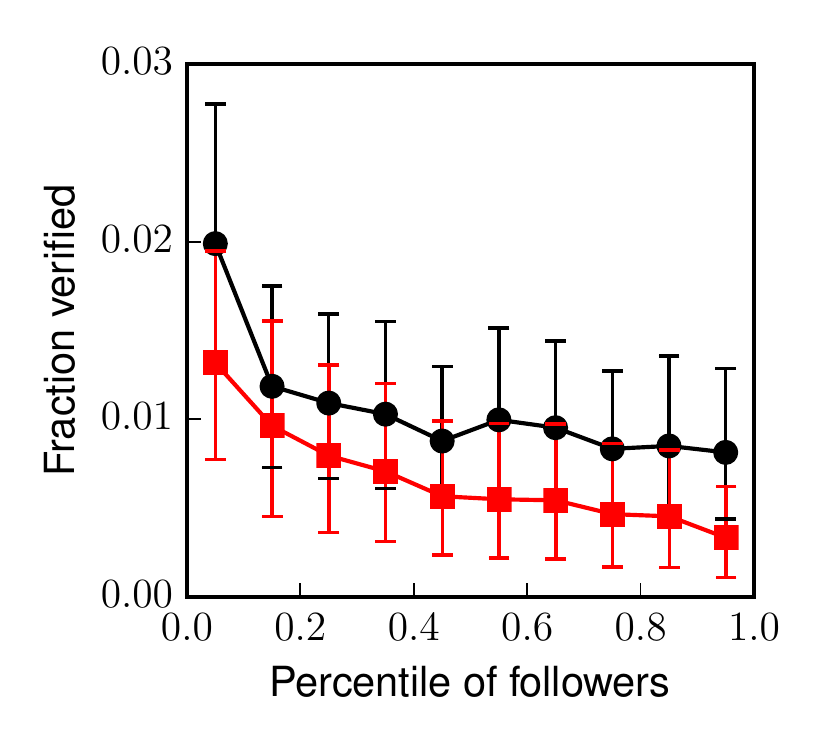}}
\subfigure[Expertise]{\includegraphics[width=0.22\textwidth]{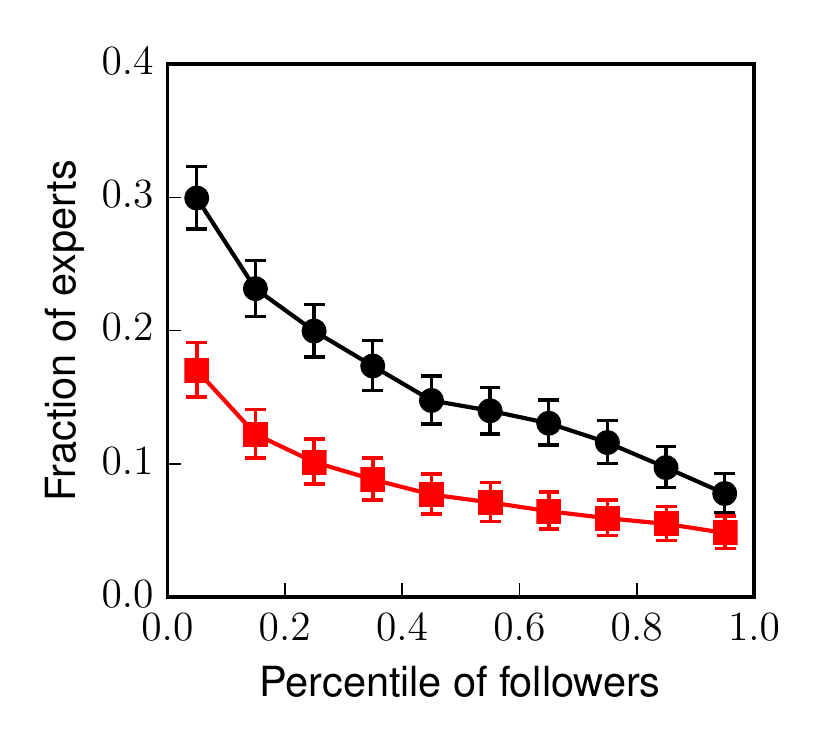}}
\caption{The eliteness of sub-groups of followers of equal size ordered by the time of link formation from the earliest to the latest for real (black circles) and reshuffled (red squares) followers. The plot is prepared for each user separately and then averaged over all users from our set of experts. The error bars are $95\%$ confidence intervals computed on the set of followers and averaged over experts; such estimate is an upper bound of the error bars. The remaining plots are prepared in the same way.}
\label{fig:eliteness}
\end{figure}

% intor
In this section, for each user from our expert set, we look at the properties of their followers as a function of the recency of these followers. This way, we measure how the early audiences differ from the later ones. 

\subsection{The eliteness of followers}
\label{sec:eliteness}

First, we quantify the eliteness of followers of our expert users. To this end, we use two different metrics of eliteness by checking whether (i) the accounts of followers are \textit{verified} by Twitter\footnote{A verification badge is granted to accounts that correspond to authentic identities of individuals or brands. More at \text{https://support.twitter.com/articles/119135}.} and (ii) whether they are collectively tagged as expert accounts~\cite{Ghosh2012Cognos:}.
% figures
We aim to measure how the eliteness of followers changes as the number of followers grows over time. 
To this end, we split the followers of each expert user into ten chronological bins: the first bin contains the $10\%$ earliest followers, whereas the last bin contains the $10\%$ latest followers. We name these chronological bins as the \textit{recency} of followers. Next, we average the measurements over followers in each bin, and then over our set of expert users.

% result
We find that the earlier followers tend to be more elite than the later followers (Figure~\ref{fig:eliteness}), both in the terms of verified and expert accounts. Later followers are around half as likely to have verified accounts and over three times less likely to be expert users (compare the leftmost and the rightmost black circles in Figure~\ref{fig:eliteness}A and~\ref{fig:eliteness}B).

\begin{figure}[t]
\centering
\includegraphics[width=0.22\textwidth]{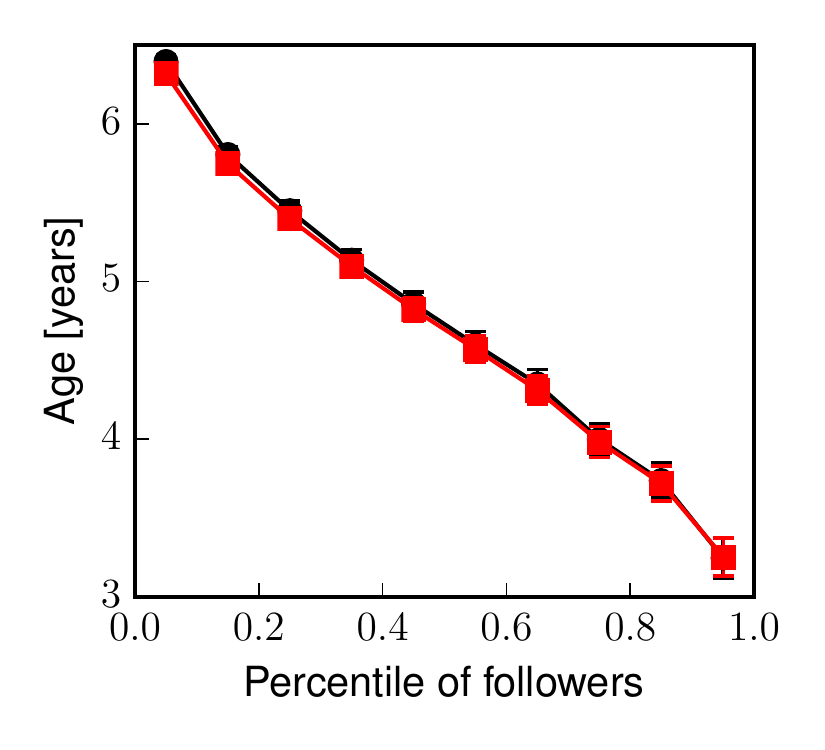}
\caption{The account's age for real (black circles) and reshuffled (red squares) followers versus their recency.}
\label{fig:age}
\end{figure}

% discussion and shuffling
We note here, that these results can be explained by at least two processes: either (i) the dilution of growing audiences of followers or (ii) the dilution of Twitter's user base itself, because later followers are users who joined Twitter more recently and soon after joining started forming their follower links (Figure~\ref{fig:age}). 
To separate the impact of these two related processes on our results, we randomly reshuffle all follower links of expert users while fixing the age of follower users in the system. Namely, for each expert, every follower is exchanged with a random follower of another expert, under the crucial constraint that an actual follower who joined Twitter in year Y can only get replaced by a random follower who also joined Twitter in year Y. Therefore, for each given expert we preserve the distribution of ages of her followers, up to variations within year Y. Furthermore, we assume that these reshuffled links for each expert were created in the same chronological sequence as the real follower links that got replaced. Thus, this reshuffling purposely preserves the correlation between age and recency of followers (compare red and black lines in Figure~\ref{fig:age}), while destroying the correlations between recency and other properties that are not implied by their correlation with age. In such a reshuffled network, existence of a correlation between recency and a given property signifies that this correlation is due to the age of followers in the whole Twitter system, instead of their recency among the given set of followers.

% result of shuffling
We plot the metrics of eliteness for the reshuffled followers as red squares in Figure~\ref{fig:eliteness}. Noticeably, the values of eliteness drop with the recency of reshuffled followers in a similar way to real followers, so the dilution of eliteness of the growing user base of Twitter is present. 
In other words, it is likely that ``tech-savy'' experts joined Twitter early on and since that time the user base of Twitter is diluting to a more casual population.
% user-level vs platform-level dilution
In the case of verified accounts, we do not find evidence for the dilution among followers of a user (curves in Figure~\ref{fig:eliteness}A are nearly parallel). However, in the case of expertise, we find that the fraction of expert users drops down faster with the recency among real followers than among reshuffled followers (the two curves meet at the rightmost point in Figure~\ref{fig:eliteness}B, but the black curve is steeper than the red curve). This finding means, that the fraction of expert users among followers drops down with the recency of these followers, providing an evidence for the dilution of expertise among growing audience at the level of followers of individual expert users.

%\subsection{Characteristics of the audience relative to the expert}
\subsection{Time zone, language, and topical congruence}

\begin{figure}[t]
\centering
\subfigure[TZ match]{\includegraphics[width=0.22\textwidth]{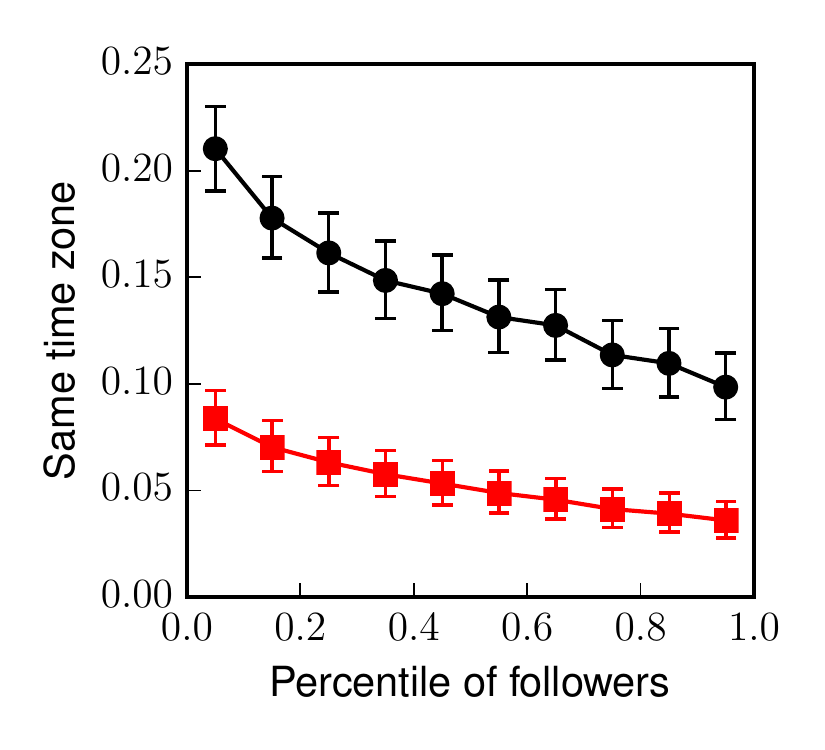}}
\subfigure[Difference in TZ offset]{\includegraphics[width=0.22\textwidth]{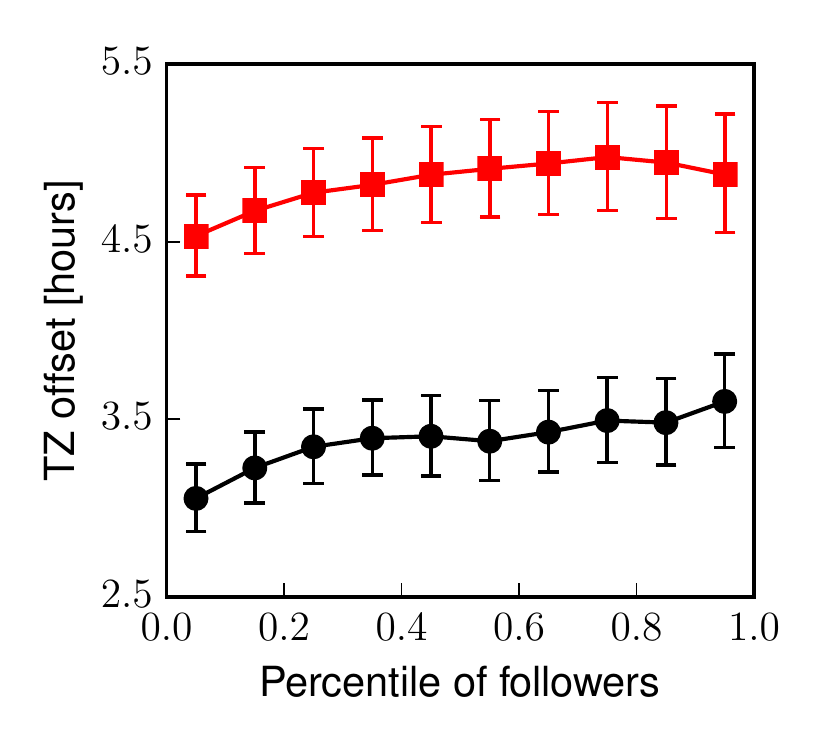}}
\caption{Time zone match between a user and her followers ordered from the earliest to the latest.}
 \label{fig:tz}
\end{figure}

\begin{figure}[t]
\centering
\subfigure[Language]{\includegraphics[width=0.22\textwidth]{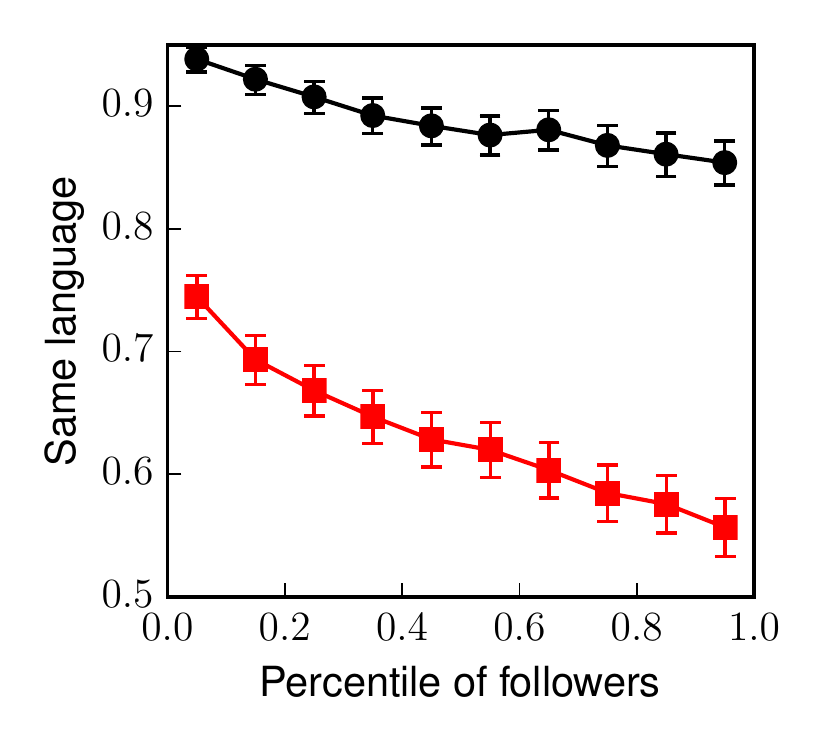}}
\subfigure[Topic]{\includegraphics[width=0.22\textwidth]{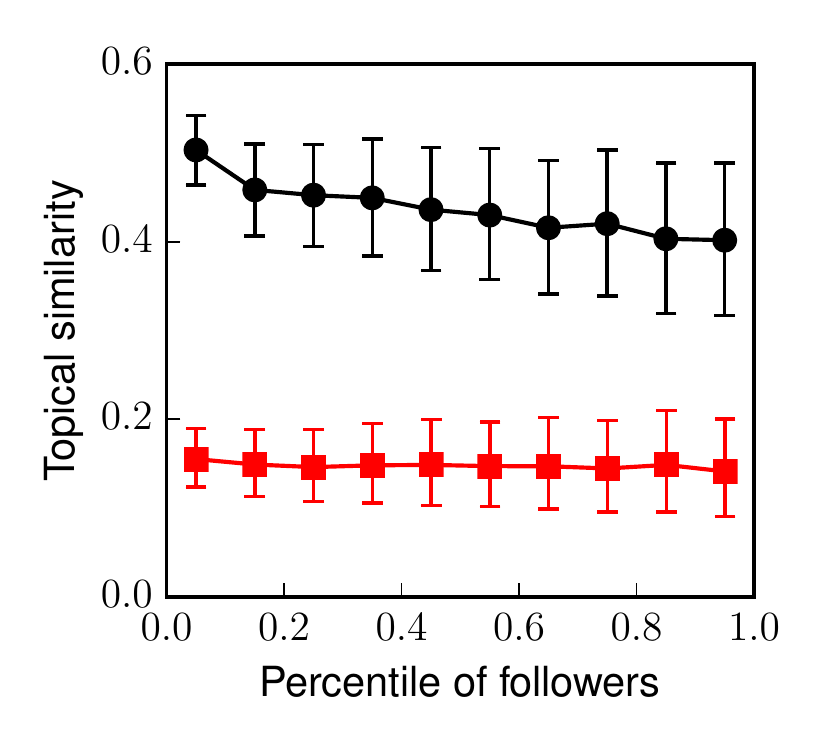}}
\caption{The language match and the topical similarity between a user and her followers ordered from the earliest to the latest.}
 \label{fig:langtopic}
\end{figure}

% relative characteristics
Next, we measure audience characteristics and compare them to the expert they follow. To this end, we extract the time zone and the language of each user from the self-reported fields specified by each user in their profile. 

% tz
Having this information, we measure if the expert and their audience are in the same time zone (Figure~\ref{fig:tz}) and if they report the same language (Figure~\ref{fig:langtopic}).
We find that the earlier followers are more likely to be in the same time zone as the person they follow (Figure~\ref{fig:tz}). The average absolute difference between the time zone of the expert and her followers is $3$ hours for the earliest followers and above $3.5$ for the latest followers; that is a nearly $20\%$ increase in the difference between the time zones. 
An increased desynchronization of time zones may lead to a coverage bias, if the experts produce certain type of content in the evening, when some of their followers are already asleep.

% language
Also, we note that the earlier followers are more likely than the later followers to use the same language as the person they follow (Figure~\ref{fig:langtopic}A). This result is followed by a related finding showing that the earlier followers are more likely to share topics of expertise with the person they follow (Figure~\ref{fig:langtopic}B). We compute the topical similarity of expertise between users as the cosine similarity of their expertise distributions~\cite{icwsm15kulshrestha}. Note that this calculation can be done only for followers who are experts themselves, for whom expertise distributions are known.

% dilution
Again, we compare these results against the results for the randomly shuffled followers. Interestingly, the dilution of language is more pronounced for the reshuffled followers. This is an intriguing result that can be explained by the fact that initially almost all tweets were in English, while nowadays there exist large non-English communities on Twitter~\cite{Liu2014Tweets}. On the other hand, we note that the topical similarity between an expert and her followers is twice larger than between the user and her reshuffled followers (compare the black and red lines in Figure~\ref{fig:langtopic}B). Most importantly, the topical alignment of Twitter's user base does not decrease, while it visibly decreases for the growing audiences of followers (compare the slopes of red and black lines in Figure~\ref{fig:langtopic}B, respectively), suggesting that the dilution is present at the level of followers of individual users, but it is not at the level of whole Twitter.

\subsection{The engagement of followers}
\label{sec:infodiff}

Finally, we would like to understand how the diluted properties of a growing audience influence engagement metrics. Among such metrics we distinguish different types of interactions between a follower and the expert that they follow, for instance: retweets, favorites, replies. Here, we focus on the results for retweets.

\begin{figure}[t]
\centering
\subfigure{\includegraphics[width=0.22\textwidth]{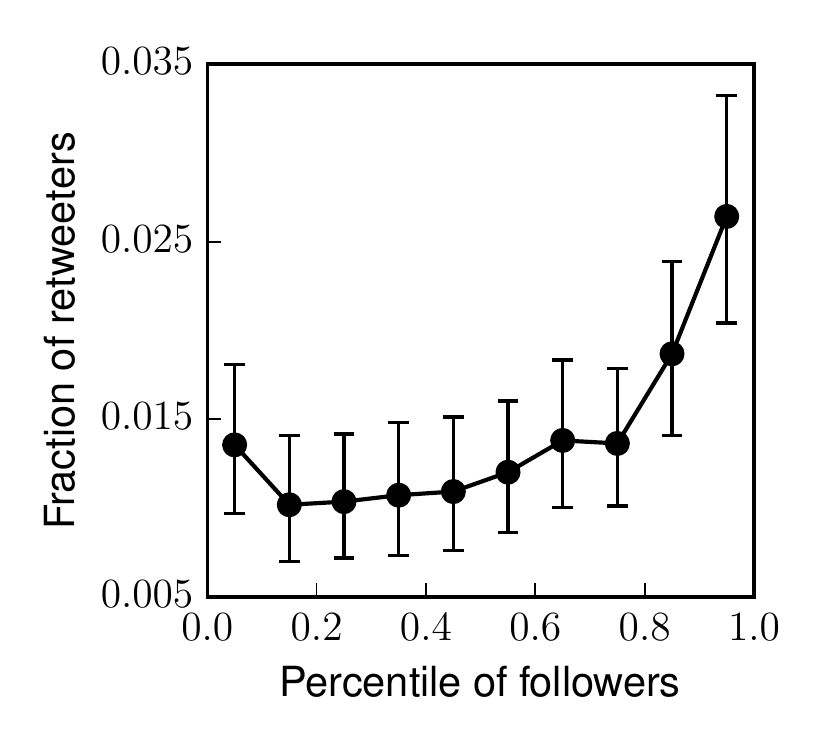}}
\subfigure{\includegraphics[width=0.22\textwidth]{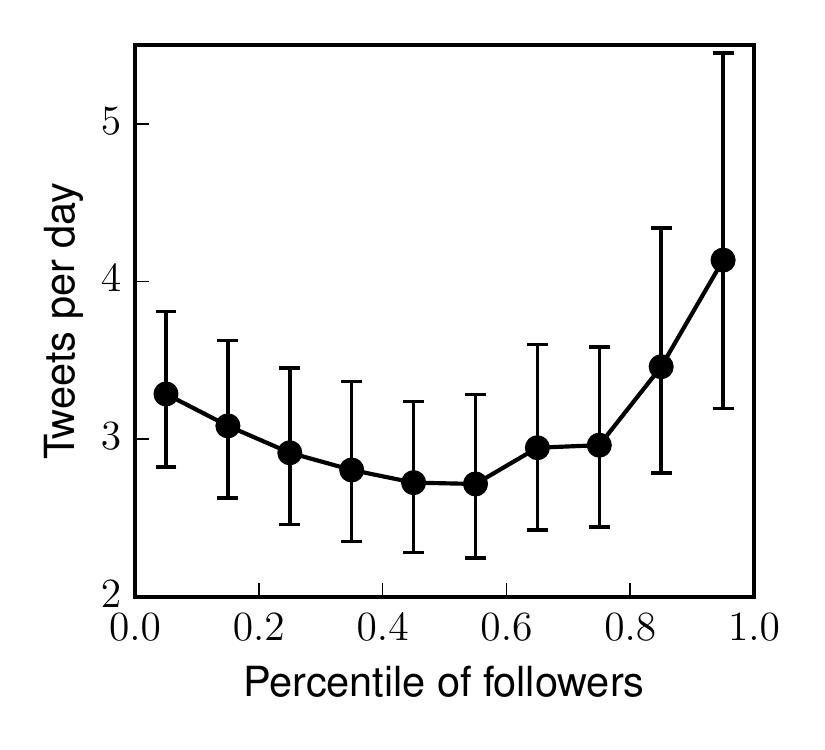}}
\caption{
(A) The fraction of retweeters among followers ordered from the earliest to the latest. 
(B) The average number of tweets produced per day by a follower versus the sequence of followers ordered from the earliest to the latest.
}
\label{fig:rts}
\end{figure}

We measure the engagement of followers with respect to the user that they follow, by measuring the fraction of retweeters among the followers of given recency (Figure~\ref{fig:rts}A).
This fraction initially drops by around $30\%$ for early followers and then it grows $3$-fold for late followers.

% interpretation
On the one hand, we expect that the fraction of retweeters among early followers should be higher, because they are more synchronized with the person that they follow in terms of time zone, language, and topics of expertise. On the other hand, the tweeting rate of the followers increases with their recency (Figure~\ref{fig:rts}B), because some users get bored and cease using Twitter, while other users are spam bots. This  may be the reason why the fraction of retweeters increases with the recency of the followers. 
%Overall, both of these explanations are plausible.

%%%%%%%%%%%%%%%%%%%%%%%%%%%%%%%%%%%%%%%%%%
\section{Discussion}
\label{sec:conclusions}

% the hypothesis of dilution of growing audiences
Our results show that the composition of growing audiences tends to get diluted. First, we find that the eliteness of followers of an expert and their topical coherence decreases with their recency. Second, we show that the composition of Twitter's user base also dilutes. For example, the eliteness and the time zone and language congruence of the users joining Twitter drops with their recency. We hypothesize that these phenomena are the effect of a common dilution process that happens for growing audiences. 

% future studies - specific
Future studies may want to understand how global evolving trends are affected and interact with the rule of dilution of growing audiences. Furthermore, it is unclear what is the impact of the growing and diluted audiences for information diffusion. We show initial measurements in this respect, but further rigorous studies of the dynamics of the latent properties of users, such as the language or topics of expertise, shall explore this question.

% recsys, limitations
Finally, we note that our measurements are platform-specific. Thus, is the law of dilution of organically growing audiences present in other systems, also offline social collectives? Furthermore, what is the importance of this law with respect to recommender systems? Oftentimes, recommender systems suggest content that is topically aligned to our interests. What minimalist recommender system leads to the dilution of growing audiences? Addressing these questions will shed more light on this intriguing phenomenon.

\bibliographystyle{aaai}
\bibliography{jabref}

\end{document}